\documentclass[12pt]{spieman}  
\usepackage{amsmath,amsfonts,amssymb}
\usepackage{graphicx}
\usepackage{setspace}
\usepackage{tocloft}
\usepackage{aas_macros}

\usepackage{caption}
\usepackage{subcaption}

\usepackage{siunitx}
\usepackage{booktabs}

\usepackage{textcomp}

\usepackage{setspace}

\usepackage{placeins}

\newcommand{\fnum}[1]{$f\!$/#1}

\newcommand{\alf}[0]{AlF$_3$}

\newcommand{\mgf}[0]{MgF$_2$}

\title{The assembly, characterization, and performance of SISTINE}

\author[a]{Nicholas Nell}
\author[a]{Kevin France}
\author[a]{Nicholas Kruczek}
\author[a]{Brian Fleming}
\author[a]{Stefan Ulrich}
\author[a]{Patrick Behr}
\author[b]{Manuel A. Quijada}
\author[b]{Javier Del Hoyo}
\author[c]{John Hennessy}
\affil[a]{University of Colorado, Laboratory for Atmospheric and Space Physics, 1234 Innovation Drive, Boulder, Colorado 80303, United States}
\affil[b]{NASA Goddard Space Flight Center, Code 551, Greenbelt, MD 20771}
\affil[c]{Jet Propulsion Laboratory, California Institute of Technology, 4800 Oak Grove Drive, Pasadena,
CA, USA 91109}

\begin{document} 
\maketitle


\begin{abstract}
  The Suborbital Imaging Spectrograph for Transition region Irradiance
  from Nearby Exoplanet host stars (SISTINE) is a rocket-borne
  ultraviolet (UV) imaging spectrograph designed to probe the
  radiation environment of nearby stars. SISTINE operates over a
  bandpass of 98 -- 127 and 130 -- 158 \si{\nm}, capturing a broad
  suite of emission lines tracing the full 10$^4$ -- 10$^5$ \si{\K}
  formation temperature range critical for reconstructing the full UV
  radiation field incident on planets orbiting solar-type
  stars. SISTINE serves as a platform for key technology developments
  for future ultraviolet observatories. SISTINE operates at moderate
  resolving power ($R\sim$1500) while providing spectral imaging over
  an angular extent of $\sim$\ang{;6;}, with $\sim$\ang{;;2}
  resolution at the slit center. The instrument is composed of an {\em
    f}/14 Cassegrain telescope that feeds a 2.1\texttimes{} magnifying
  spectrograph, utilizing a blazed holographically ruled diffraction
  grating and a powered fold mirror. Spectra are captured on a large
  format microchannel plate (MCP) detector consisting of two 113
  \texttimes{} 42 mm segments each read out by a cross delay-line
  anode. Several novel technologies are employed in SISTINE to advance
  their technical maturity in support of future NASA UV/optical
  astronomy missions. These include enhanced aluminum lithium fluoride
  coatings (eLiF), atomic layer deposition (ALD) protective optical
  coatings, and ALD processed large format MCPs. SISTINE was launched
  a total of three times with two of the three launches successfully
  observing targets Procyon A and $\alpha$ Centauri A and B.
\end{abstract}


\section{Introduction}
\label{sect:intro}  

SISTINE is a rocket-borne far-ultraviolet (FUV; 91.2 -- 180 \si{\nm})
spectrograph previously described in Refs. \citenum{France2016} and
\citenum{Fleming2016}. This paper focuses on the implementation of
SISTINE and describes assembly, characterization, and laboratory and
flight performance of the instrument. Technology development efforts
and results for SISTINE are discussed. Characterization of the
instrument and the facilities that allow complete in-band
characterization of the instrument are described. SISTINE was launched
a total of three times, two flights of which successfully acquired
target data. The final flight of SISTINE notably launched from the
Arnhem Space Center in the Northern Territory, Australia, allowing
observation of a Southern target, $\alpha$ Centauri A+B. Descriptions
of modifications and improvements made to the instrument between
flights are summarized. We describe the final flight performance of
the instrument, demonstrating the capabilities of SISTINE to probe the
FUV radiation environments of nearby stars and mature key optical
technologies now planned for adoption by larger NASA astrophysics
missions.

\section{SISTINE Science Objectives}

Characterization of exoplanet atmospheres, including the potential for
habitability, requires an understanding of the interaction with the
host star's UV radiation environment \cite{Hu_2012, France_2013,
  Lammer_2003}. F, G, K, and M type dwarf stars (approximately 20 -
0.1 solar masses) are hosts to Earth-mass planets located in habitable
zones. Many of these systems will be searched for biomarkers by NASA's
large missions including the James Webb Space Telescope (JWST)
and the upcoming Habitable Worlds Observatory (HWO). FUV radiation
impacts chemicals such as H$_2$O, CO$_2$, and CH$_4$. The host star's
Ly$\alpha$ is a significant contributor to these effects and can also
be used as a proxy for the extreme-ultraviolet (EUV; 10 -- 91.2
\si{\nm}) flux from the host star \cite{Youngblood_2016_a}. EUV flux
is likely to have a significant effect on atmospheric mass loss on
exoplanets but is difficult to measure directly due to attenuation by
the interstellar medium (ISM). Additionally, flare activity in the FUV
can be correlated to large ejections of charged particles that can
heavily affect O$_3$ quantity in an atmosphere
\cite{Segura_2010}. SISTINE's imaging capability and spectral
resolution allow the investigation of low-mass star FUV environments
and their effects on potential exoplanet atmospheres. The spatial axis
imaging capability of SISTINE allows the subtraction of geocoronal
Ly$\alpha$. With the ability to subtract the geocoronal contribution,
the targeted spectral resolution of SISTINE (R $\sim$7000 at 121.6
\si{\nm}) is intended to resolve and reconstruct the target star's
Ly$\alpha$ emission line profile\cite{Youngblood_2016_a,
  Youngblood_2022}. SISTINE provides spectral coverage from 98 -- 127
and 130 -- 158 \si{\nm}. This range spans strong atomic emission lines
tracing various formation temperatures in the stellar atmosphere:
$10^{4}$ K (Ly$\alpha$, 121 nm), $10^5$ K (C IV, 155 nm), and
$10^{5.5}$ K (O VI, 103 nm).

\section{Instrument Description}

\subsection{Instrument Design}

\begin{figure}[ht]
  \centering
  \includegraphics[width=\textwidth]{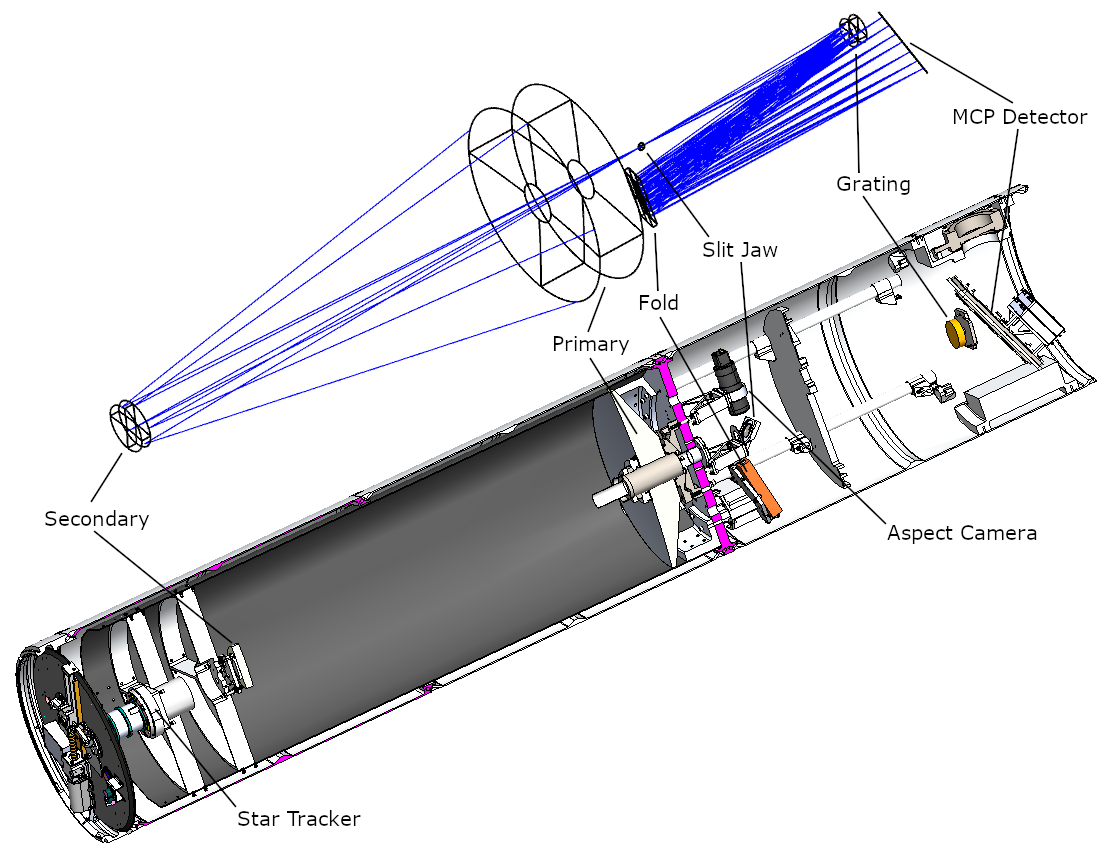}
  \caption{\label{fig:raytrace} The SISTINE optical raytrace (upper)
    and mechanical CAD (lower) layouts are shown. Light enters the
    system on the left side of the image. The star tracker, used to
    acquire attitude and maintain low jitter during flight, is located
    in the telescope aperture and coupled to the optical metering
    structure. The aspect camera that images the slit jaw and allows
    fine pointing maneuvers in flight is shown near the center of the
    system. The spectrograph MCP detector is located on the right side
    of the image and is the focal plane responsible for capturing
    spectra.}
\end{figure}

The design of SISTINE has been modified since the descriptions in
previous literature and we present the final design of SISTINE as
fabricated. SISTINE is an imaging spectrograph. Target light is imaged
by an \fnum{14} Cassegrain telescope with a 0.5 \si{\m} diameter
primary mirror onto a \ang{;;10} \texttimes{} \ang{;6;} entrance
slit. The spectrograph is composed of a concave spherical
holographically ruled grating, a concave cylindrical fold mirror, and
a microchannel plate (MCP) detector. This optomechanical layout is
shown in Figure \ref{fig:raytrace}. Target light passes through the
entrance slit, is dispersed by the grating, reflected off of the fold
mirror, and captured on the detector. The cylindrical figure of the
fold mirror serves to match the tangential and sagittal foci of the
grating on the focal plane, as well as reduce the linear dispersion at
the focal plane maintaining the desired bandpass on the detector.

The slit jaw is angled \ang{45} relative to the telescope
boresight. Light that does not pass through the entrance slit is
reflected off of the 36 \si{\milli\meter} diameter polished aperture
surface onto an aspect camera. The aspect camera field of view is
approximately \ang{;12.5;} $\times$ \ang{;17.5;}. The video feed from
the aspect camera is monitored in real time to align the instrument to
targets in flight or calibration environments. The aspect camera in
SISTINE is a Photonis Nocturn XL that contains a CMOS sensor optimized
for low light levels without the use of an intensifier. Commercially
available optical lenses are assembled in a ruggedized lens tube to
image the slit jaw focal plane onto the camera sensor.

\begin{figure}[ht]
  \centering
  \includegraphics[]{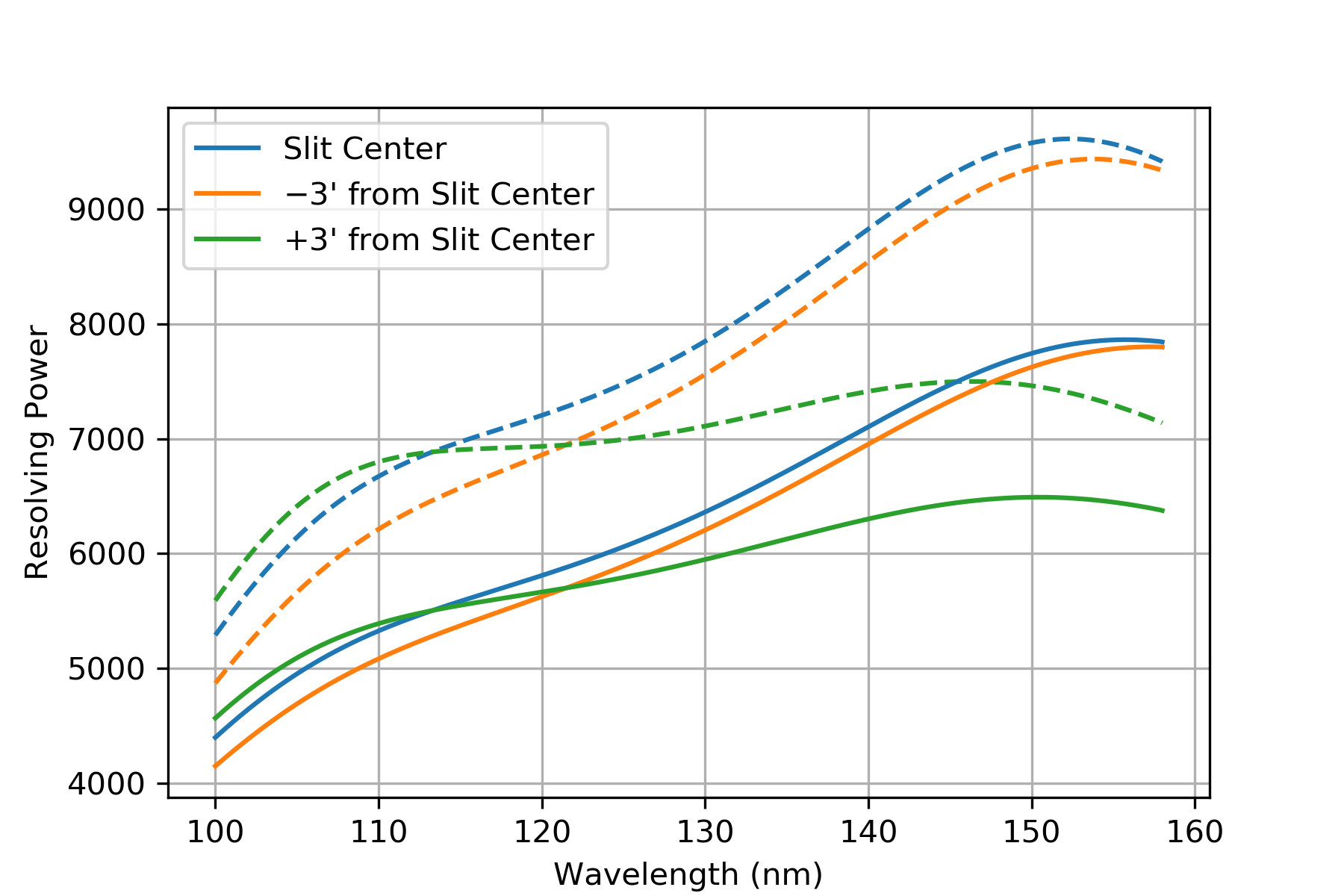}
  \caption{\label{fig:resolvingp} Modeled SISTINE resolving power for
    a telescope design with a \ang{;;0.6} FWHM PSF. Dashed lines
    correspond to a detector resolution element size of 40 \si{\um}
    (specified) and solid lines correspond to a detector resolution
    element size of 62 \si{\um} (empirical). Resolving power is shown
    at three points across the imaging axis, at the center of the slit
    and at either end of the slit.}
\end{figure}

Light that passes through the slit is incident upon the grating. The
SISTINE grating is holographically ruled for a central density of
3016.6 g/\si{\mm} and blazed via ion etching. The blaze angle of
\ang{10.9} is optimized to maximize efficiency in the -1 order at a
wavelength of 120 \si{\nm}. The grating is mounted with a $4.6 \pm
0.1$ \textdegree tilt such that zero order light is directed towards a
light trap and light from the -1 order is directed towards the fold
mirror and reflected onto the detector. Details on the grating, fold,
and detector are provided in Table \ref{tab:inst}. This design results
in an average dispersion of 0.256 \si{\nm}/\si{\mm} at the
detector. Because the detector is composed of two separate segments
there is a small unimaged gap in the center of the spectrum resulting
in a bandpass of 98 -- 127 and 130 -- 158 \si{\nm}. The spectrograph
optical system is $\sim$ \fnum{29.4} and thus magnifies the point
spread function (PSF) imaged from the telescope focal plane by a
factor of $\sim$ 2.1 on the detector. This results in a PSF FWHM size
incident upon the detector of $\sim$ 40 \si{\um} (without factoring in
spectrograph aberrations). The specified detector resolution element
(40 \si{\um} FWHM) is matched to this PSF and results in a minimum
detectable PSF FWHM size of $\sim$ 63 \si{\um} when spectrograph
aberrations are factored in, resulting in a resolving power of
$\sim$ 7400 at 120 \si{\nm}. Predicted resolving power for a
\ang{;;0.6} FWHM telescope PSF is shown in Figure
\ref{fig:resolvingp} for detector resolution element sizes of 40
\si{\um} (specified) and 62 \si{\um} (empirical).

SISTINE employs a large format MCP detector designed and fabricated by
Sensor Sciences LLC. It is composed of two independent segments each
with an active area of 113 $\times$ 42 \si{\mm}. A gap of $\sim 11.7$
\si{\mm} exists between the two segments where light is not
detected. Each segment is a ``Z'' format stack of three borosilicate
atomic layer deposition (ALD) processed microchannel plates with 20
\si{\um} pores and a \ang{13;;} bias angle. The pore bias angle is
aligned along the instrument cross dispersion axis. An opaque cesium
iodide (CsI) photocathode was deposited on the upper plate of each
segment. Each segment is read out by a cross delay line (XDL) anode
with anode delays of $\sim 69$ \si{\ns} in the X axis and $\sim 49$
\si{\ns} in the Y axis. Digitization in both spatial axes is comprised
of 14 bits, pulse height is digitized with 10 bits with a full photon
position and pulse height encoded in 38 bits.

\begin{table}[ht]
\caption{SISTINE instrument design summary.} 
\label{tab:inst}
\begin{center}       
  \begin{tabular}{l l}
    Telescope & \\
    \toprule
    Geometric Collecting Area & 1830 cm$^2$ \\
    Focal Length & 7000 \si{\mm} \\
    Field of View & \ang{;12.5;} $\times$ \ang{;17.5;} \\
    Primary Diameter & 500 \si{\mm} \\
    Primary Radius & 3000 \si{\mm} \\
    Primary Conic & $-1.0$ \\
    Secondary Diameter & 120 \si{\mm} \\
    Secondary Radius & 810.22 \si{\mm} \\
    Secondary Conic & $-2.388$ \\
    \hline
    Spectrograph & \\
    \toprule
    Bandpass & 98 -- 127, 130 -- 158 \si{\nm} \\
    Spectral Resolving Power & $\sim 2000$ \\
    Entrance Slit & \ang{;;10} \texttimes{} \ang{;6;} \\
    Grating Diameter & 75 \si{\mm}\\
    Grating Radius & 850.39 \si{\mm} \\
    Grating Ruling Density & 3016.6 g/\si{\mm} \\
    Grating Blaze Angle & \ang{10.9} \\
    Fold Dimensions & 165 \texttimes{} 50 \si{\mm} \\
    Fold Radius & 7000 \si{\mm} \\
    \hline
    Detector & \\
    \toprule
    Active Segment Size & 113 \texttimes{} 42 \si{\mm} \\
    Spectral Axis Resolution Element & 62 \si{\um} \\
    Pore Size & 20 \si{\um} \\
    Pore Bias Angle & \ang{13} \\
    \hline

  \end{tabular}
\end{center}
\end{table}

\begin{figure}[ht]
  \centering
  \includegraphics[width=\textwidth]{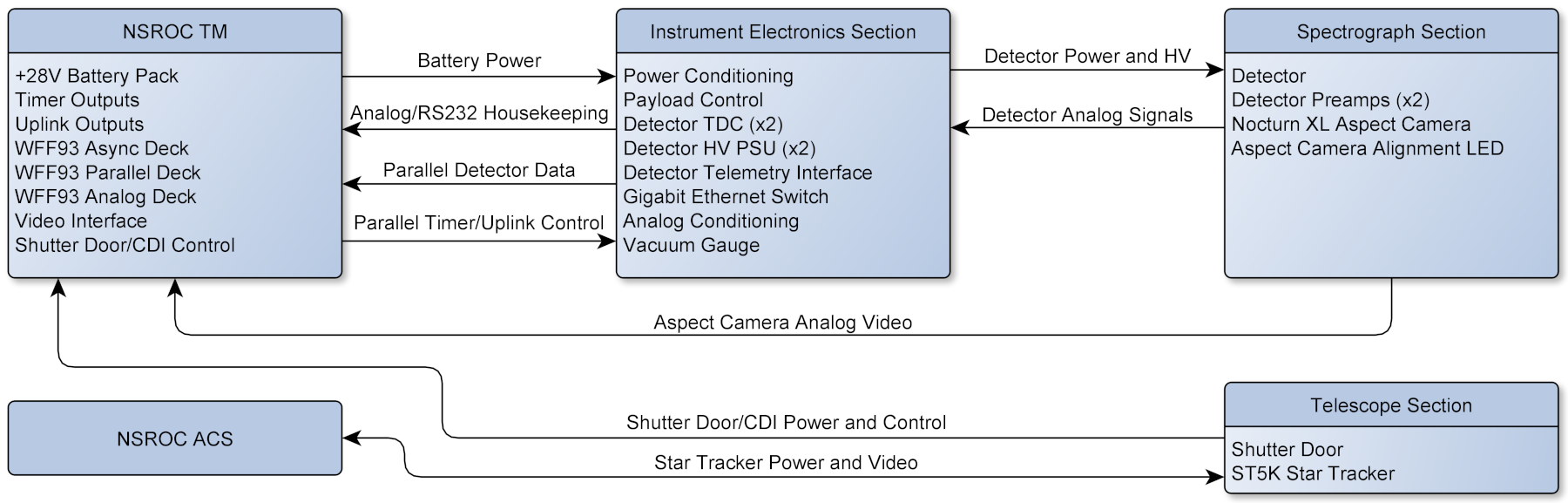}
  \caption{\label{fig:bd} Electrical system block diagram of the
    SISTINE payload. Instrument electronics are hermetically isolated
    from the telescope and spectrograph sections. Critical data paths
    are shown as directional connections between the various blocks.}
\end{figure}

The instrument electronics section contains the readout electronics
for the detector, high voltage power supplies to bias the detector
MCPs, instrument control electronics, analog signal conditioning, a
vacuum gauge, and a telemetry interface to format detector data for
downlink. A block diagram of these subsystems is shown in Figure
\ref{fig:bd}. A power conditioning board powers all of these
subsystems from a single 28 V battery pack located in the telemetry
section built by the NASA Sounding Rocket Operations Contract (NSROC)
engineers. An instrument control board reads uplink and timer signals
from the telemetry section (TM) and allows real time control of the
high voltage (HV) biasing, aspect camera alignment LED, and reset of
the telemetry interface (TMIF). The control board also monitors
payload pressure and instrument status and passes that data to the TM
for downlink. Thermistor voltages, instrument voltages, and instrument
currents are conditioned by the analog signal conditioning board and
passed to a WFF93 analog deck that digitizes those values and
telemeters them for real time monitoring during flight and
testing. Analog detector data from each segment is amplified by a pair
of preamplifiers located close to the detector in the spectrograph
section. The amplified signals are transmitted to a pair of time to
digital converters (TDC) with the end of each anode signal being
passed through an appropriate delay line to match the anode
delay\cite{Vallerga2000}. The TDCs digitize these analog signals and
output ethernet packets containing X position, Y position, and a pulse
heigh for each photon. Ethernet from each TDC is routed to a
ruggedized commercial gigabit ethernet switch and data from the switch
is routed to the TMIF. The TMIF parses all of this data, performs
minor encoding, and outputs data in a synchronous parallel format to a
WFF93 parallel deck in the TM section. This system enables real time
monitoring of detector data during testing and flight.

\subsection{Technology Development}

SISTINE is used as a platform for the maturation of novel technologies
relevant to astrophysical applications in the FUV. These technologies
are ALD processed borosilicate MCPs with large formats, enhanced
lithium fluoride (eLiF) optical coatings tailored to the FUV, and
environmentally protective ALD optical coatings. Borosilicate MCPs
allow larger format detectors to be built and ALD processing of the
MCPs allows control of gain and MCP resistance. While Al+LiF mirrors
have been used successfully on several past missions the eLiF effort
focuses on improving the quality of the LiF layer by optimizing the
temperature of the substrate. Finally, efforts to protect the LiF
layer with a final thin ``capping'' layer of a protective material
with sufficient transmissivity is carried out via ALD. These efforts
are discussed in detail in the following two sections.

\subsubsection{Large format MCP detectors}
\label{sec_mcp}

\begin{figure}[ht]
  \centering
  \includegraphics[width=\textwidth]{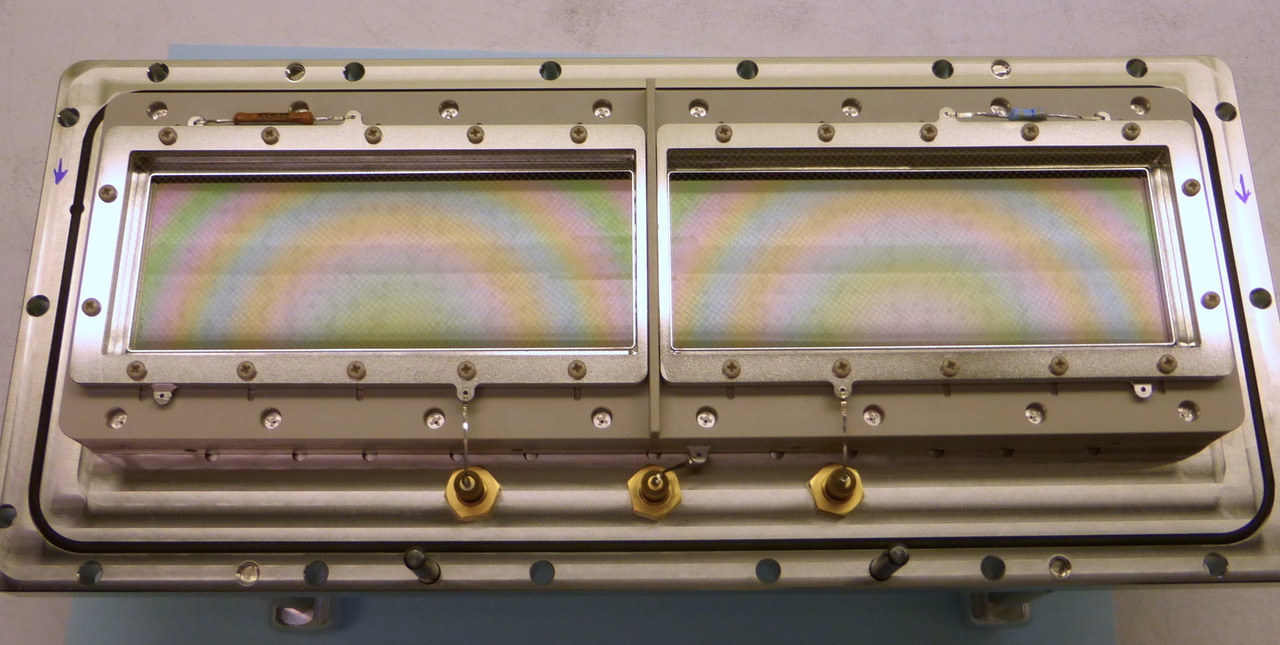}
  \caption{\label{fig:det} The fully assembled SISTINE MCP
    detector. The shorter wavelengths fall on the left segment and
    longer wavelengths fall on the right segment. Each segment is 113
    \texttimes{} 42 \si{mm}. Minor thickness variations in the CsI
    photocathode are responsible for the fringing effect visible on
    each segment. A QE enhancement grid is visible oriented in a
    rotation roughly \ang{45;;} off of the primary segment axes.}
\end{figure}

ALD processed borosilicate MCPs are available in formats up to 200
\texttimes{} 200 mm and these types of plates are shown to have better
gain stability than traditional plates\cite{Siegmund_2021,
  Siegmund_2020, Ertley_2018, Ertley_2015, Popecki_2016}. This work
builds upon the state of the art detectors flown on FUSE and HST-COS
which both used segment sizes of 85 \texttimes{} 10 mm and each
employed two segments. SISTINE has segment sizes of 113 \texttimes{}
42 mm and also employs two segments with an XDL readout for each
segment. Resolution elements of \SI{62}{\micro\meter} along the X axis
and \SI{38.5}{\micro\meter} along the Y axis were achieved with the
SISTINE detector characterized via pinhole mask testing performed at
Sensor Sciences LLC. The SISTINE detector with photocathode applied is
shown in Figure \ref{fig:det}. An opaque CsI photocathode was applied
to the top surface of each MCP stack, resulting in detector efficiency
performance comparable to that of HST-COS\cite{Vallerga2001} and
GOLD\cite{Siegmund2016} and efficiency measurements are provided in
Section \ref{sec:comp_performance}.

The SISTINE detector employs recent advances in MCP technology, ALD
processed MCPs made of borosilicate substrates. A notable aspect of
this detector is the large active area, a total of 94.92 \si{cm}$^2$,
made possible by advances in MCP fabrication technology. The MCPs for
the SISTINE detector were fabricated by Incom, who have pioneered the
development of the borosilicate MCP
substrate\cite{Popecki_2016}. These MCPs have resistance and secondary
emission characteristics driven by ALD. The ALD processed borosilicate
MCPs allow for the fabrication of large formats with improved gain
stability and lifetime while maintaining low background
rates\cite{Siegmund_2021, Siegmund_2020, Ertley_2018,
  Ertley_2015}. Only basic functional testing was performed at the
Laboratory for Atmospheric and Space Physics (LASP) prior to detector
installation in SISTINE. Throughout the course of SISTINE testing and
flight, no notable changes were observed in detector behavior. This
detector was refurbished and new photocathodes were deposited between
the first and second SISTINE flights in September, 2020.

\subsubsection{Enhanced protective optical coatings for the FUV}
\label{sec_coatings}
  
Efforts to improve the reflectivity of aluminum coated optics in the
FUV have been employed on multiple optics in SISTINE through the use
of eLiF and an ALD protective coating\cite{Fleming2015,
  Fleming2017}. While Al+LiF has been deposited and flown successfully
on FUSE it is known that Al+LiF mirrors are sensitive to humidity and
can suffer reflectivity degradation over time when exposed to
environments with sufficient humidity\cite{Oliveira_1999}. The eLiF
technology development effort has focused on improving reflectivity in
the FUV and simultaneously improving the environmental stability of
the coated mirrors. The eLiF effort employs elevated substrate
temperatures with the goal of lowering the extinction coefficient of
the LiF layer. Evaporated LiF has been shown to have a significantly
higher extinction coefficient than crystalline LiF at short
wavelengths which is undesirable for high reflectivity
coatings\cite{Dauer2000}. In addition, the ALD technology development
focuses on protecting the LiF layer, deposited via physical vapor
deposition, with a final layer of another more environmentally
resistant material deposited via ALD\cite{Hennessy2016}. Three optics
in SISTINE were coated with Al+eLiF, the primary mirror, secondary
mirror, and fold mirror. The secondary mirror was also coated with a
capping layer of \alf{} via ALD\cite{Hennessy2016}. We note that the
grating flown on SISTINE is a replica and thus was not able to be
coated with eLiF due to the material properties of the replica not
being compatible with the substrate temperatures required for eLiF
deposition. Replica gratings are produced using an epoxy layer on the
replica substrate to generate a high fidelity copy of the originally
fabricated master grating\cite{Loewen_1983}. Discussions with the
grating vendor, Horiba Jobin-Yvon, concluded that the epoxy used in
their replication process is not compatible with the high substrate
temperatures required for eLiF deposition. Deposition of eLiF is
performed via physical vapor deposition but the optical substrate is
brought up to a temperature of approximately 250 \si{\celsius} prior
to deposition to improve packing density of the LiF, thus reducing
porosity and surface roughness. All optical substrates for SISTINE
were radiatively heated with infrared lamps with a proxy sample used
to estimate temperature of the optical substrate. The secondary and
fold mirrors both showed improved reflectivity while the primary
mirror did not show notable improvement over ambient temperature LiF
deposition. The primary mirror has a significantly larger volume than
the secondary or fold mirrors making it extremely difficult to heat
uniformly at high temperatures in a vacuum environment to achieve the
desired substrate temperature for deposition\cite{deMarcos_2022}. We
note that it was not possible to measure the SISTINE primary mirror
directly due to the size and mass of that mirror so all results for
the primary mirror are inferred from the witness sample which was
closely thermally coupled to the primary mirror via a hub mount
located in the central cutout of the primary mirror during the coating
process. Details of handling and storage of the optics and witness
samples is explained in Section \ref{sec:comp_performance}.

\begin{figure}[ht]
  \centering
  \includegraphics[]{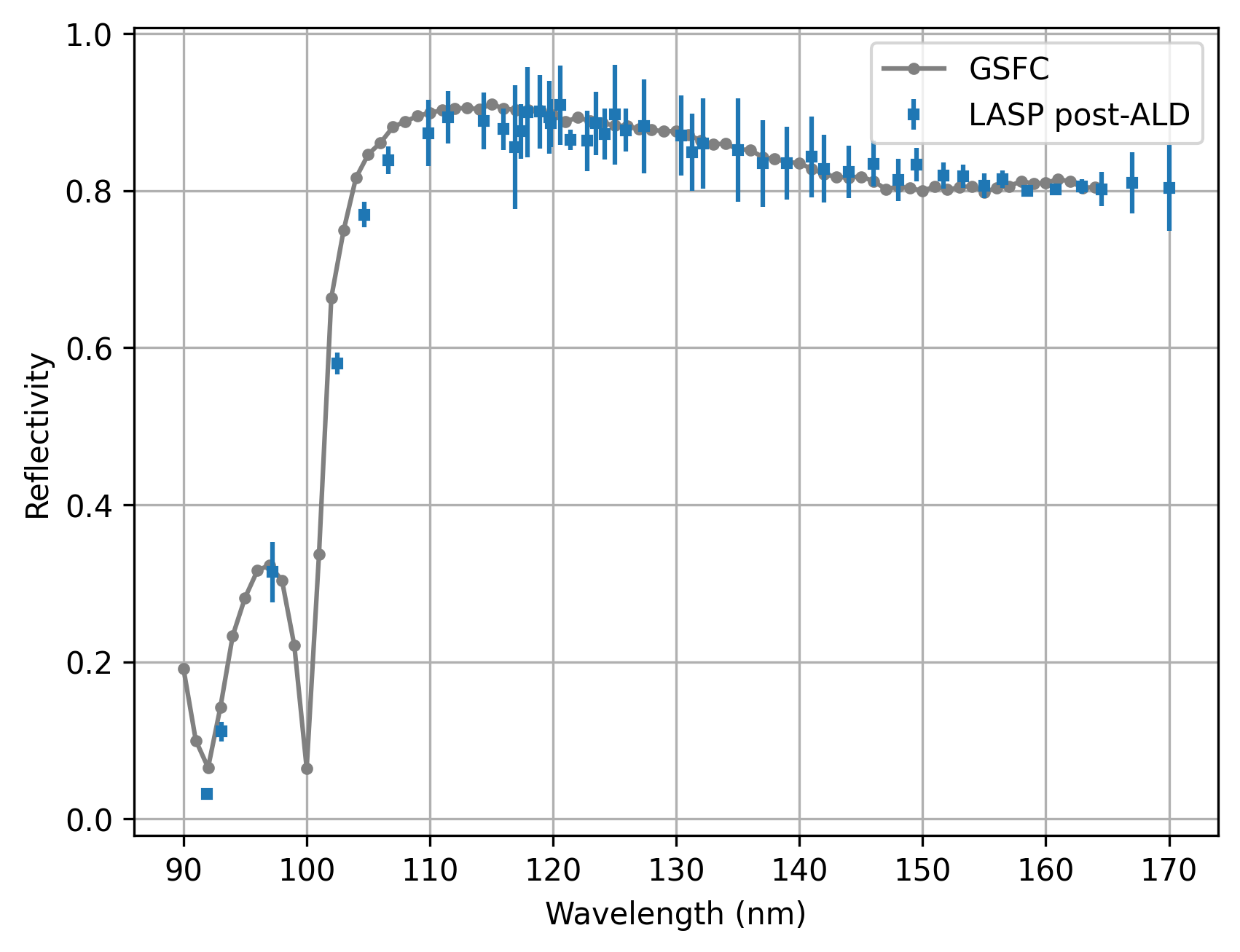}
  \caption{\label{fig:secwitness} Reflectivity measurements of the
    SISTINE secondary mirror witness sample measured after deposition
    of Al+eLiF at GSFC (grey points) and after ALD \alf{} coating
    (blue points). Data shown in grey are measurements performed by
    GSFC and data shown in blue are measurements performed by
    LASP. The \alf{} ALD coating has only a minor effect on the
    overall reflectivity of the sample most pronounced at the LiF
    cutoff wavelengths near 102 \si{nm}. This sample was the best
    demonstration of the capability of eLiF with peak reflectivities
    of 90\% near 115 \si{nm}.}
\end{figure}

\begin{figure}[!htb]
  \centering
  \includegraphics[]{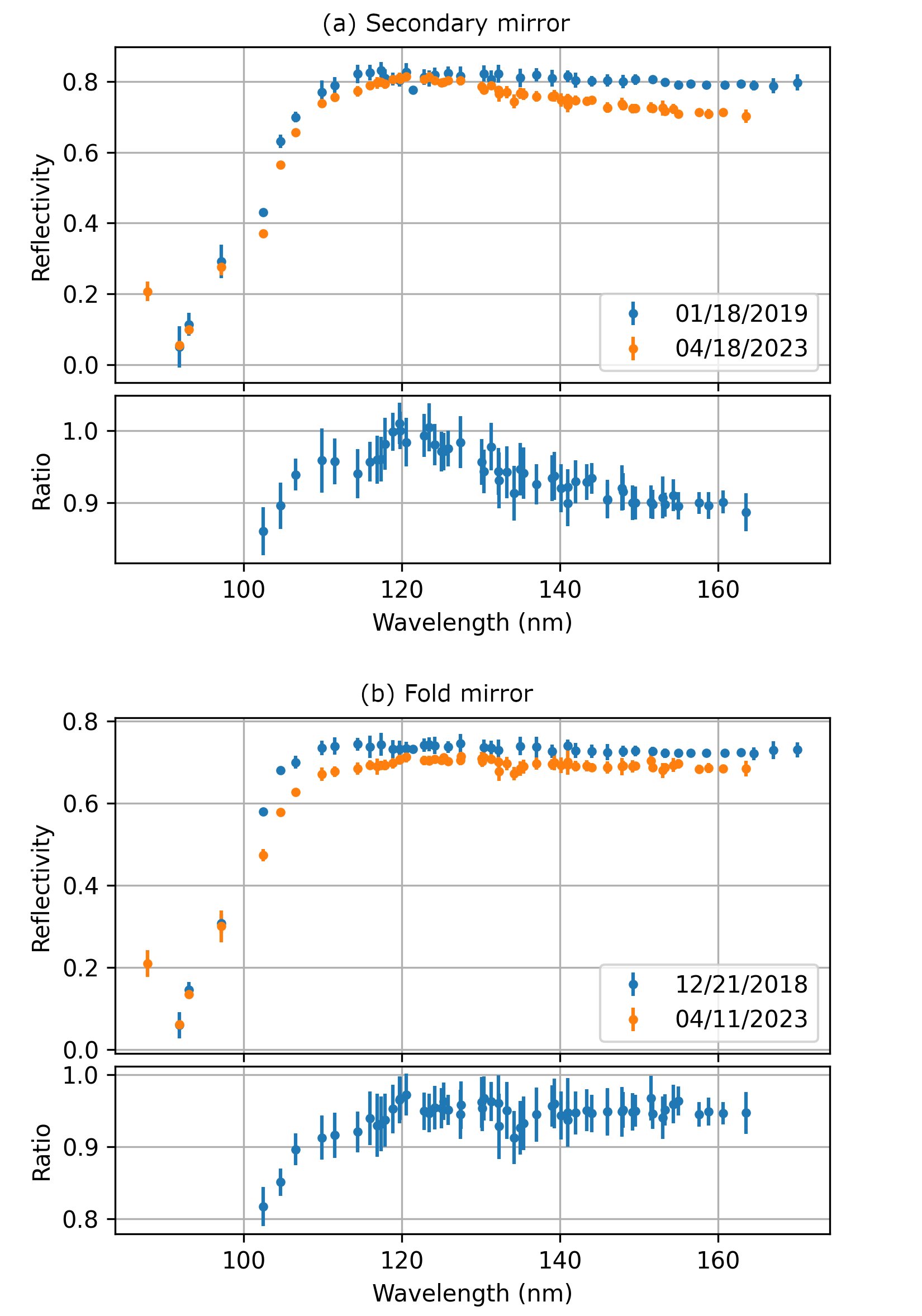}
  \caption{\label{fig:secfold} Reflectivity measurements for the
    secondary (a) and fold (b) mirrors from before installation into
    the payload and after the final flight of SISTINE are shown. The
    lower panels show the relative change from initial to final
    measurements. These measurements provide an empirical instrument
    usage case demonstration of performance over time for these two
    coating prescriptions.}
\end{figure}

It is also notable that witness samples for the secondary and fold
mirrors had higher reflectivities than the optics themselves. These
results implied that one of the challenges with achieving good eLiF
results is controlling substrate temperature. Ultimately the best
results achieved were on the secondary witness sample, shown in Figure
\ref{fig:secwitness}, which demonstrate the potential of eLiF with
peak reflectivities of 90\% near 115 \si{\nm}. The efficiencies
observed suggest that the eLiF deposition process reduces the
extinction coefficient of the LiF layer at short wavelengths but a
measurement of the indices of refraction for eLiF has not yet been
made. The secondary mirror and secondary witness sample were
additionally coated with 10 cycles of \alf{} via ALD at Jet Propulsion
Laboratory to protect the LiF layer. The \alf{} deposition had a
minimal effect on the overall reflectivity (see Figure
\ref{fig:secwitness}). No controlled aging studies were possible with
the SISTINE optics but we show a comparison of the secondary and fold
optics measured prior to installation into the SISTINE instrument and
then after the three launches in Figure \ref{fig:secfold}. Both optics
maintained high reflectivity throughout their active usage in SISTINE
and show different aging trends. Note that the aging trend seen for
the secondary mirror shown in Figure \ref{fig:secfold} is
qualitatively similar to the change in reflectivity seen after ambient
aging for an Al+\alf{} mirror sample shown in Reference
\citenum{Quijada_2017}. This appears to show that the ALD \alf{} layer
is serving the intended purpose and is the material primarily
interacting with the environment. These results suggest that \alf{} in
particular may be of interest when stability at the shortest
wavelengths is paramount. The ratio values shown in the lower panels
of Figure \ref{fig:secfold} are only shown down to the cutoff
wavelength because the extinction coefficient of evaporated LiF shows
little difference to that of crystalline LiF at wavelengths short of
the cutoff\cite{Dauer2000}. Efforts to develop high efficiency LiF
coatings have continued after the SISTINE results and are discussed in
Refs. \citenum{deMarcos_2022} and \citenum{Quijada_2022}.

\subsection{Mechanical Design}

\begin{figure}[ht]
  \centering
  \includegraphics[]{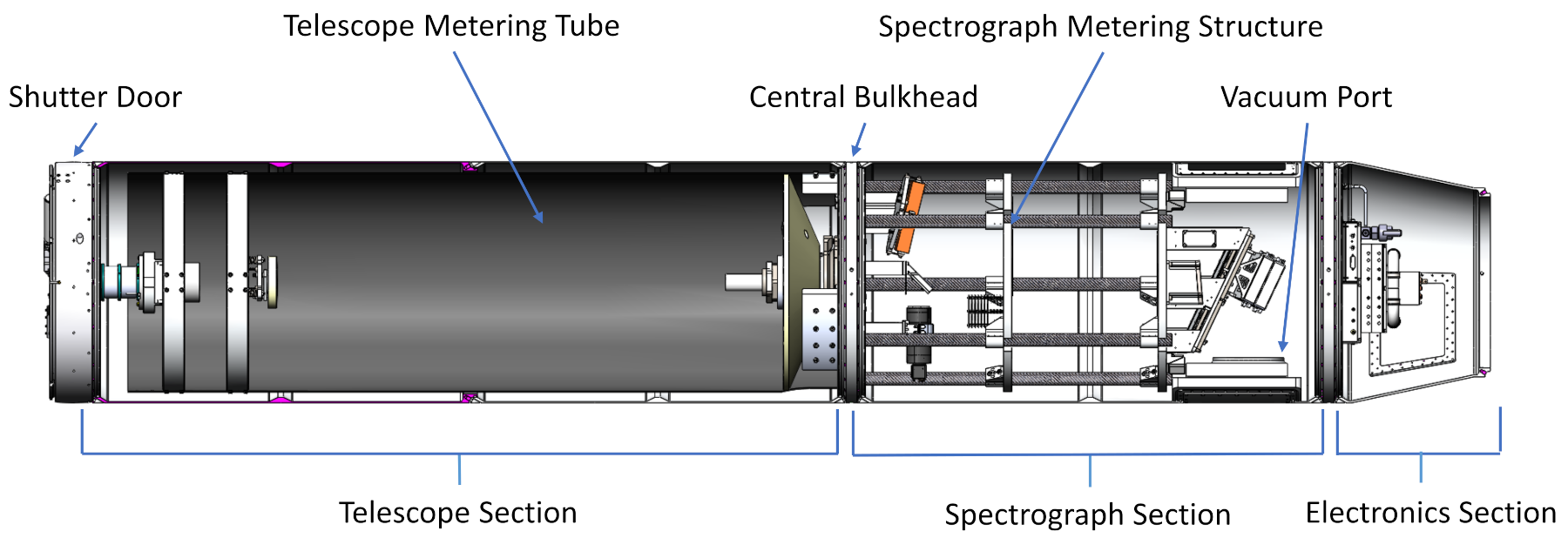}
  \caption{\label{fig:mech} Mechanical layout of the SISTINE
    instrument. The telescope and spectrograph sections are a single
    hermetic manifold. All optomechanical structures are metered off
    of the central bulkhead. Optical metering for both the telescope
    and spectrograph is accomplished with carbon fiber parts. A large
    vacuum pump port is located in the spectrograph section allowing
    the entire system to reach high vacuum for testing in the
    laboratory.}
\end{figure}

SISTINE is an aft-looking payload housed in 22'' hermetic aluminum
skins. There are three distinct mechanical sections, the telescope,
the spectrograph, and the electronics (see Figure \ref{fig:mech}). The
telescope and spectrograph sections are a single hermetic manifold
that can be pumped through a valved port to achieve low pressures
($<1\times{}10^{-5}$ Torr) where the detector can be operated for
testing. The hermetic section also allows the payload to be filled
with nitrogen to keep the optics and detector safe during shipment or
down time when the payload cannot be actively pumped. The telescope
aperture is closed with a hermetic shutter door designed and
fabricated by NSROC. The electronics section is built on a bulkhead
that seals the other end of the telescope and spectrograph hermetic
manifold.

The optical system is metered off of a single central bulkhead located
between the telescope and spectrograph sections (see Figure
\ref{fig:mech}). This mounting scheme attempts to mitigate thermal
effects from the rapid heating of the outer aluminum skins during
ascent through the atmosphere. The telescope and spectrograph metering
sections are independent fabrications and employ different designs
suited to each application. The spectrograph metering structure is an
open frame with aluminum decks for optical mounts. This structure
allows access to spectrograph optical components to facilitate
optical alignment.

\section{Instrument Assembly and Characterization}

\subsection{Characterization Facilities} \label{sec:characterization_facilities}

\begin{figure}[ht]
  \centering
  \includegraphics[width=\textwidth]{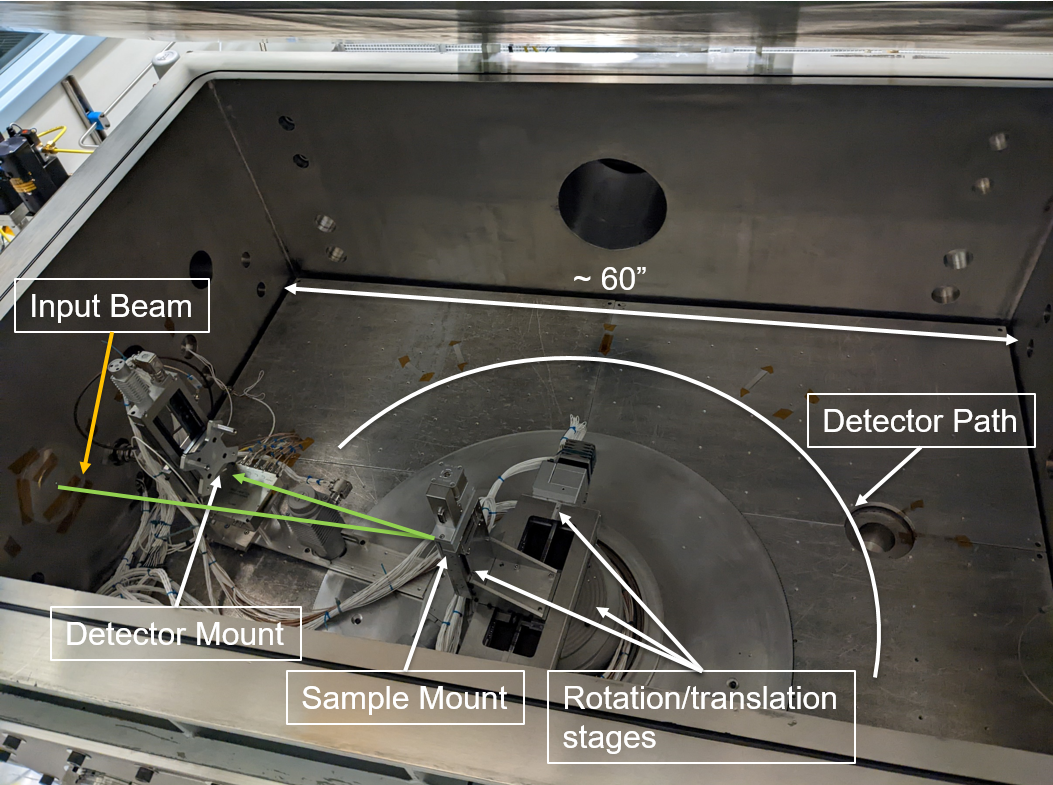}
  \caption{\label{fig:sqtank} Annotated photograph of the square tank
    showing sample manipulation stages, sample and detector mounts,
    input beam path, and detector path. The configuration shown is
    typical for reflectivity measurements with the detector shown in
    the reflected beam measurement position. The incident beam is
    measured by the same detector which travels along the detector
    path with the sample moved out of the optical path via the sample
    manipulation stages. This system allows reflectivity to be
    measured by a single detector. A similar configuration is used for
    grating efficiency testing.}
\end{figure}

Component and instrument characterization in-band occurs in two
facilities at LASP. Component characterization is performed in the
``square tank'', originally described in Ref. \citenum{Windt1986} and
used to measure optical constants for a wide variety of coating
materials in the EUV and FUV\cite{Windt1988a, Windt1988b}. Since that
time the facility has undergone some changes and a more recent
description is contained in Ref. \citenum{Kruczek2022}. The chamber
and sample manipulation stages retain nearly identical functionality
and have some modularity to be used for normal or grazing incidence
measurements as well as having the capability to accommodate different
grating measurements. Two monochromators are used to illuminate
samples, a McPherson 310 is used for wavelengths spanning 1 -- 121.6
\si{nm} and an Acton VM-502 is used for wavelengths spanning 90 -- 200
\si{nm}. It is possible to make measurements all the way through 550
\si{nm} but in practice this is much less efficient than using
commercial spectrophotometers that can measure down to $\sim185$
\si{nm}. Three light sources are commonly used for measurements in
this chamber, a Manson Model 2 soft x-ray source, a windowless hollow
cathode discharge lamp\cite{Paresce1971}, and a Hamamatsu L9841
deuterium lamp with an \mgf{} window. Two detectors are commonly used
to measure across this full bandpass. An open faced MCP detector with
an XDL anode and opaque CsI photocathode from Sensor Sciences LLC is
used to measure shorter wavelengths. This detector uses a hybrid ALD
and traditional MCP stack to achieve high gain stability despite
repeated illumination in a small area. For longer wavelengths a
Hamamatsu R6836 photomultiplier tube (PMT) is used. This PMT has an
\mgf{} window and a CsTe photocathode and is used for measurements of
wavelengths spanning 115 -- 300 \si{nm}. An annotated photograph of
this facility is shown in Figure \ref{fig:sqtank}.

End-to-end instrument characterization is performed in the ``long
tank'' chamber, originally described in Ref. \citenum{Cook1991}. The
``long tank'' is a 23 foot long cylindrical vacuum chamber with a 30
inch inner diameter accommodating an entire sounding rocket payload
within the vacuum manifold. One end of the tank contains a Newtonian
collimator with a 24 inch {\em f}/4 mirror. The collimator optics are
coated with gold and illuminated using a windowless hollow cathode
discharge lamp mounted on a stage on the side of the chamber allowing
fine field position and focus adjustment. The angular size of the
source is set with an interchangeable pinhole or mask at the focal
plane of the collimator allowing different field illuminations to be
used including a series of point sources (see Section
\ref{sec:spec_cal}). The instrument is optically aligned to the tank
in tip and tilt using a manual stage near the chamber door. Fine
adjustments are then made using the light source adjustment stage to
sample different positions across the slit.

\subsection{Instrument Component Performance}
\label{sec:comp_performance}

\begin{figure}[ht]
  \centering
  \includegraphics[]{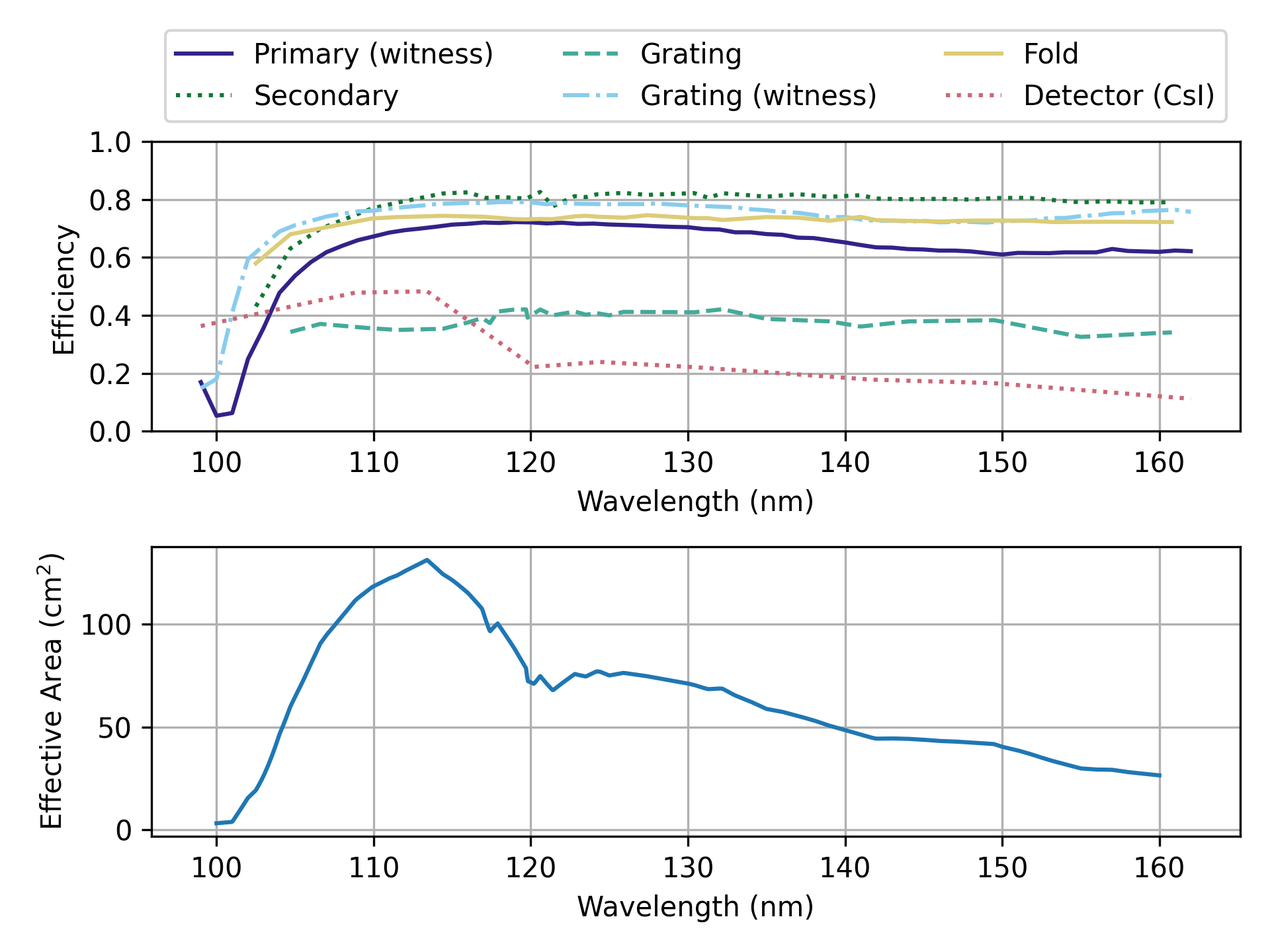}
  \caption{\label{fig:aeff} (Top) Efficiencies for all optical
    components and the detector are shown. (Bottom) The effective area
    of SISTINE derived from the efficiencies of the individual
    component is shown as a function of wavelength. Peak predicted
    effective area is $\sim$ 120 \si{\cm}$^2$ near 115 \si{\nm}.}
\end{figure}

The expected performance of the SISTINE instrument is modeled from
measurements of the various components of the system. Predictions for
SISTINE efficiencies and effective area were published in 2016
\cite{Fleming2016}. Coating prescriptions are discussed in Section
\ref{sec_coatings} and the detector photocathode and measurements are
discussed in Section \ref{sec_mcp}. All of the SISTINE optics and all
witness samples underwent final characterization in the LASP square
tank facility, except for the primary mirror which is too large to be
tested in the square tank. The estimate for the reflectivity of the
primary mirror is derived from the witness sample that was coated with
the primary mirror. The flight grating efficiency was tested in the
same geometric configuration as used in SISTINE but a $\sim$ 1
\si{\cm} circular portion near the center of the grating was
illuminated for this testing. For flight optics only one measurement
position is shown, but several measurements are taken across the
surface to verify consistent performance across the optical
surface. Results from these tests are shown in Figure \ref{fig:aeff}
and are used to estimate the effective area shown in the lower panel
of Figure \ref{fig:aeff}.

All optics and associated witness samples were transported in hermetic
containers backfilled with N$_2$ to maintain a low humidity
environment during transport. Prior to installation in the payload all
optics were stored in an active N$_2$ purge cabinet in a clean
room. During all processes witness samples are maintained in close
proximity with their corresponding optics to track any potential
issues during such activities and no efficiency issues were observed
throughout all bonding procedures. In general, the witness samples for
each optics are maintained in close proximity with the associated
optic until final installation into the payload. When optics were
installed into the payload their corresponding witness samples were
left in N$_2$ active purge storage which maintains a low humidity
environment. Handling of the optics only occurs in clean workspaces
with care taken to avoid any types of molecular contamination of the
optics or assemblies. Optics in the payload are occasionally exposed
to the ambient atmosphere during buildup and testing
operations. However, the payload itself is hermetic and generally kept
at high vacuum when possible. During shipment operations the payload
is backfilled with N$_2$ to a pressure of $\sim$ 200 Torr.

\subsection{Telescope Assembly and Characterization} \label{sec:tscope}

The entire optical system is metered off of the primary bulkhead (see
Figure \ref{fig:mech}). The primary and secondary mirrors of the
telescope were fabricated by Nu-Tek Precision Optics. The primary
mirror is mounted directly to the primary bulkhead on a tip/tilt
mount. The primary mirror is fixed in place to the mount with a
threaded clamp around the inner diameter of the central cutout of the
mirror. This scheme was chosen due to the tight space constraints
imposed by the mirror size and lightweighting scheme versus the
available volume in a 22 inch sounding rocket manifold. After this
mirror is installed, the carbon fiber (CF) telescope metering tube is
attached to the primary bulkhead with large aluminum brackets. This
metering tube was designed at LASP and manufactured for SISTINE by
Rock West Composites. A camera is temporarily attached to the metering
tube at the focal point of the SISTINE primary mirror. This subsystem
is then fully illuminated with collimated light and the primary mirror
is adjusted in tip and tilt to minimize the size of the PSF and remove
aberrations such as coma. Once the PSF is optimized the primary mirror
adjustment mechanism is locked and staked with low outgassing epoxy.

The camera is then removed from the metering tube and the secondary
optic is installed on its support structure. An autocollimation test
fixture is placed at the nominal telescope focal plane and the
telescope is pointed at a flat reference mirror. The secondary optic
is adjusted in tip, tilt, and piston to minimize the size of the PSF
at the focal plane. The focal plane can be sampled both directly with
a camera and indirectly with a knife edge test to measure PSF
size. After the secondary mirror position is optimized, lock nuts and
set screws are used to firmly hold it in place and all adjustment
hardware is staked. The autocollimation test fixture is then removed
and replaced with the slit jaw assembly and the aspect camera is
installed.

\begin{figure}[ht]
  \centering
  \includegraphics[width=\textwidth]{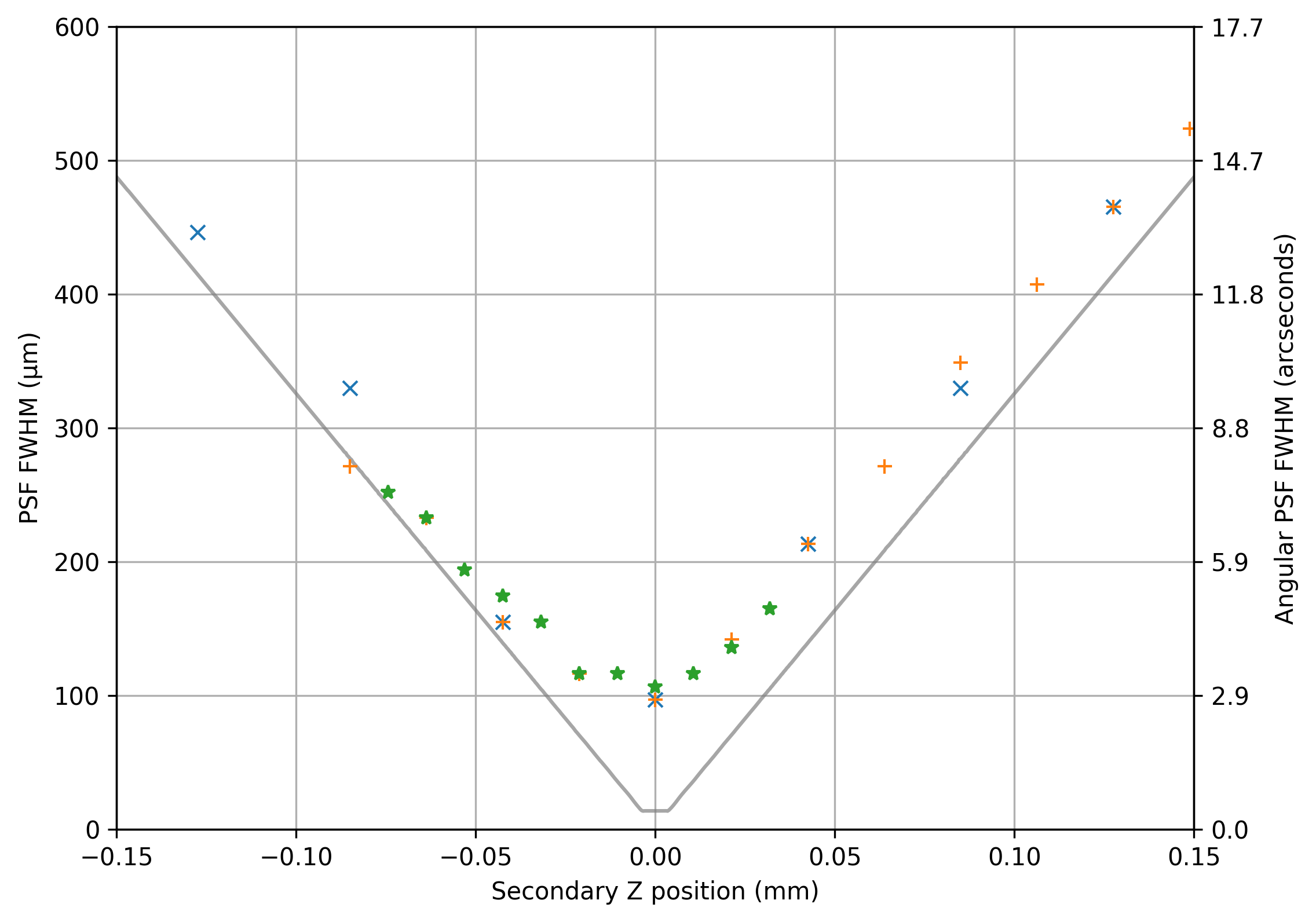}
  \caption{\label{fig:telefocus} A double pass focus curve as a
    function of the secondary mirror position for the SISTINE
    telescope is shown. Grey lines show the theoretical focus curve as
    predicted with Zemax OpticStudio. Points in blue, orange, and
    green show measured PSF FWHM size using different step sizes.}
\end{figure}

Telescope focus prior to the first flight of SISTINE is shown in
Figure \ref{fig:telefocus}. This focus measurement was accomplished
using a light source with a wavelength of $550$ \si{nm} in air. The
minimum value reached is $\sim$ 100 \si{\um} FWHM which is
$\sim$\ang{;;3} FWHM in angular space. We note that this measured
value is the result of a double pass test where the wavefront error is
doubled. The single pass PSF that would be measured in flight could
therefore be as small as $\sim$\ang{;;2} FWHM in flight. This PSF was
larger than the originally targeted value of
\ang{;;0.6}\cite{Fleming2016} and was expected to have been caused by
stress induced on the primary mirror from the mount. There was no time
to investigate this issue during the assembly of SISTINE 1. During the
SISTINE 1 flight, the mirror was found to have come loose in its
mount, no spectral data of NGC 6826 was acquired during that
flight. There was no damage to the mirror, so the mount was modified
to double the clamping torque applied and a lock nut was added to the
primary clamping nut for future flights. The telescope was again
focused in a similar manner before the flight of SISTINE 2 and results
of that focus run were consistent with SISTINE 1 overall but more
features were apparent in the PSF. This is believed to be due to
stress induced deformation in the primary mirror from the need to
increase the clamp torque. The secondary support spider was initially
made of multiple segments but after the first flight a monolithic
spider was fabricated to ensure no change in secondary position under
flight vibration environment.

\begin{figure}[ht]
  \centering
  \includegraphics[width=\textwidth]{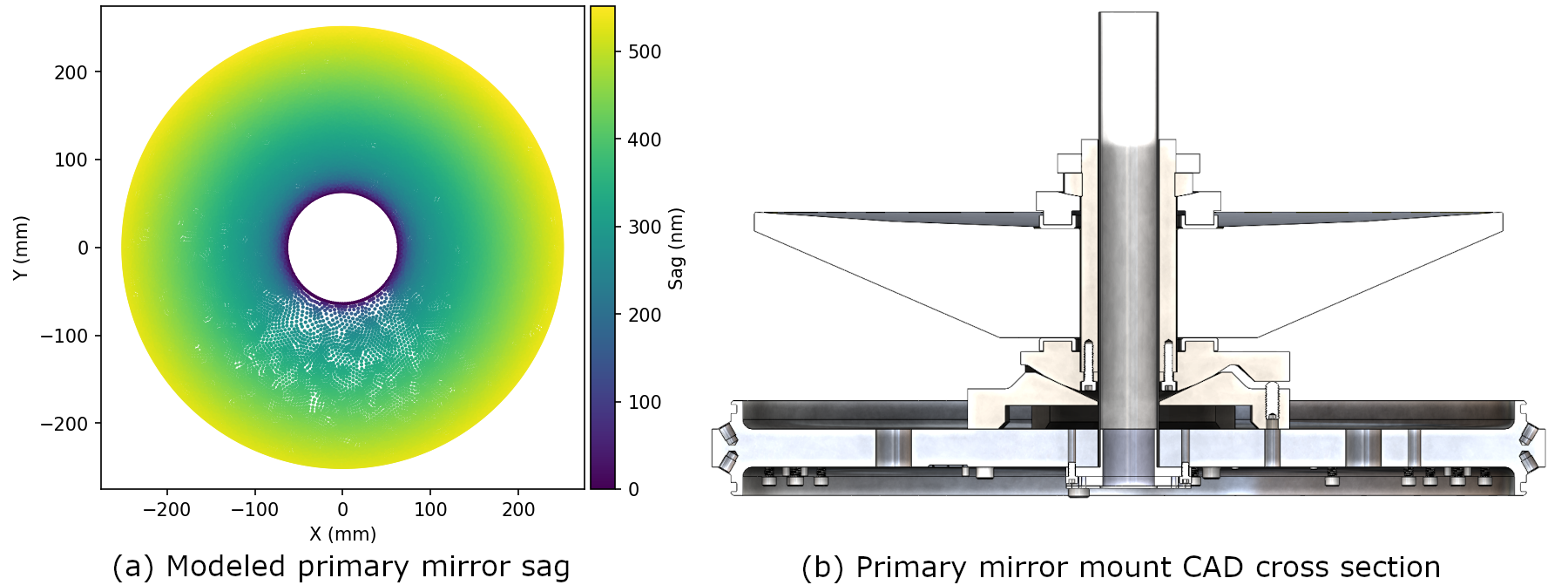}
  \caption{\label{fig:telesag} (a): FEA modeled primary mirror
    deformation from the ideal figure with a clamping torque of 154
    foot pounds. (b): CAD sectional view of the SISTINE primary mirror
    hub mount.}
\end{figure}

This deformation of the primary mirror was modeled using finite
element analysis (FEA) with the simplified assumption that the torque
was uniform azimuthally across the clamping surface against the optic
(see Figure \ref{fig:telesag}). Even though this analysis does not
perfectly model the stress on the optic it provides the magnitude of
the figure error introduced from this level of torque. The sag change
from the FEA analysis was fit with Zernike polynomials and this system
was modeled in Zemax OpticStudio which provided results consistent
with our measurements. This work occurred in late 2019 and there was
not sufficient time to redesign the primary mirror mount for future
flights of SISTINE. In late 2019 the SISTINE team was working towards
a July 15, 2020 launch date in Australia. Payload integration and
shipment schedules only allowed a few months for instrument updates
and those efforts were ultimately focused on preventing the issue
experienced on the SISTINE-1 launch. In early 2020 the COVID pandemic
and condition of the Australia site caused an indefinite postponement
of the SISTINE-2 launch date.

Between the SISTINE 2 and 3 flights, a point source microscope (PSM)
similar to the design shown in Refs. \citenum{Steel1983} and
\citenum{Parks2005} was fabricated and characterized for use in
optical alignments. During the flight of SISTINE 2 it was observed
that the telescope was out of focus and that the PSF produced by the
telescope was larger than the width of the slit
(\ang{;;10})\cite{CruzAguirre2023}. However, when the system was
tested after flight, no defocus was observed. A series of tests were
conducted to find the root cause of the defocus in flight, including
gravity loading and thermal environment. Ultimately the thermal
environment during the flight was found to have caused the
defocus. The SISTINE telescope was separated from the rest of the
system and placed on an optical bench in a double pass configuration
with the PSM mounted on a linear stage at the focal plane such that
the focus could be tracked quantitatively in real time. Industrial
heat tape was applied to the exterior of the telescope skin and
controlled with a variac to reproduce temperatures on the bulkhead as
observed during flight. During the heating process the focus was
observed to change as a function of bulkhead temperature. Once flight
like temperatures were reached the distance the focus had moved in the
optical axis was measured to be 3.8 \si{\mm}, but at this new location
the size of the PSF was still consistent with the nominal focus on
SISTINE 1 and 2 as observed in the laboratory. Using the observed
thermal 3.8 \si{\mm} shift to predict the geometric size of the PSF
results in a prediction of $\sim$ \ang{;;27} which is consistent with
the PSF size observed in flight for SISTINE
2\cite{CruzAguirre2023}. The primary mirror is the optic in the
telescope optical path with the most thermal coupling to the central
bulkhead which experiences a significant temperature increase ($\sim$
15 \si{\celsius}) during vehicle ascent.  Again, there was no time to
redesign a primary mirror mount and the schedule for SISTINE 3
laboratory work was abbreviated due to the constraints of testing and
shipment for an international campaign. Therefore, to mitigate this
defocus effect for flight, the telescope was purposefully defocused by
3.8 \si{mm} in the opposite direction of the empirically determined
thermally induced focus shift by adjustment of the piston position of
the secondary mirror. The system was thermally tested again and the
results of this solution were found to be consistent with previous
laboratory focus results. SISTINE 3 successfully acquired data from
$\alpha$ Cen A+B demonstrating an in-flight PSF consistent with that
measured in the laboratory\cite{Behr_2023} demonstrating that the
thermal characterization and executed solution operated as intended
(see Section \ref{sec:flight_performance}).

\subsection{Spectrograph Assembly and Characterization} \label{sec:spec_cal}

The spectrograph is built on the opposite side of the primary bulkhead
from the telescope. A space frame was built to meter the spectrograph
grating and detector. The design is based on the space frame from the
Colorado High-resolution Echelle Stellar Spectrograph which had been
flight proven and shown to be well suited to the sounding rocket
environment\cite{Hoadley2014, Kruczek2018}. The grating was fabricated
by Horiba JY and the fold mirror was fabricated by Rainbow Research
Optics. The fold mirror and aspect camera are mounted directly to the
primary bulkhead. The only adjustable optic in the spectrograph is the
fold mirror which has tip, tilt, and piston adjustment to properly
align the spectrum on the detector and bring it into focus. Three
motorized linear actuators are used to adjust the fold mirror so that
the spectrograph can be focused in real time while the entire payload
is in a vacuum chamber and illuminated with collimated FUV light. Once
the fold position is optimized, it is locked and staked into place and
the linear actuators are removed.

\begin{figure}[ht]
  \centering
  \includegraphics[width=\textwidth]{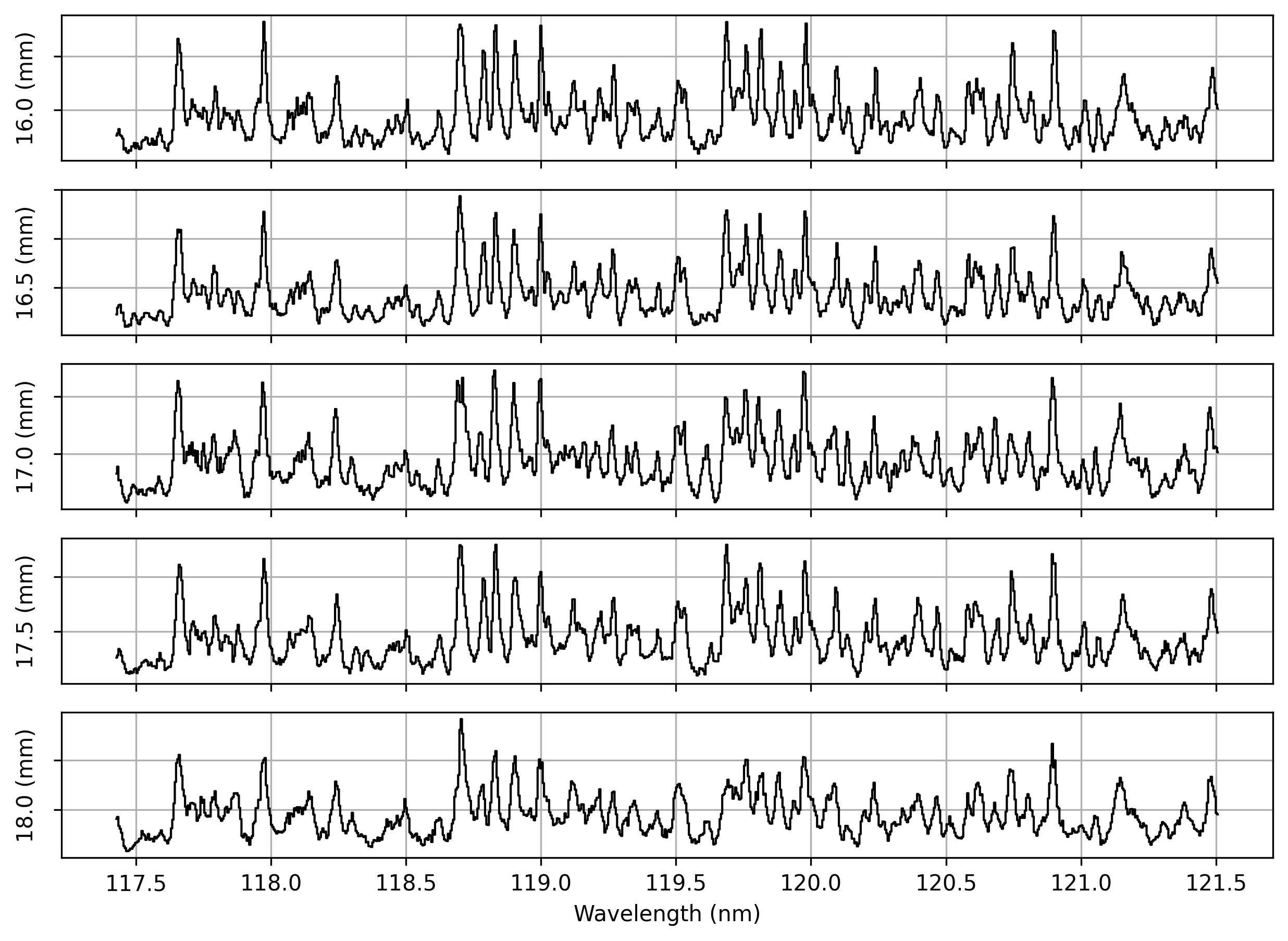}
  \caption{\label{fig:specfocus} A set of deuterium spectra taken at
    different fold mirror positions. The Y axis labels show position
    of the fold mirror in mm along the optical axis. The intensity of
    each spectrum is shown in arbitrary units. The wavelength solution
    is approximate as each spectrum has to be positionally corrected
    after fold mirror positioning in post processing for
    comparison. Although data is taken over the full bandpass only a
    small subset is shown to demonstrate detail.}
\end{figure}

In the SISTINE design, the slit is oversized and does not drive
spectral resolving power of the system. This is a feature meant to
ensure that all of the flux from the observed source is captured to
maximize sensitivity of the system. However, due to the performance of
the telescope, it was not possible to demonstrate the full capability
of the spectrograph. For focusing and characterization, a pinhole
aperture was placed at the location of the slit to effectively
decrease the size of the source being fed into the spectrograph. This
aperture was 35 \si{\um} \texttimes{} 50 \si{\um} with the 35 \si{\um}
portion aligned along the spectral axis. Factoring in the plate scale
of the telescope, this source has an angular size of \ang{;;1}
\texttimes{} \ang{;;1.5}. All spectrograph adjustments were made with
this aperture in place. Focus operations were possible with a
windowless hollow cathode source but there are a limited number of
lines that are useful for characterizing limiting performance of the
spectrograph itself. The N I 124.3 \si{nm} line was used for a coarse
focus adjustment followed by a fine focus adjustment using a deuterium
lamp and a similar method to the ``shotgun'' approach described in
Ref. \citenum{Wilkinson1998} for FUSE. Figure \ref{fig:specfocus}
shows a small subset of deuterium spectra near optimal focus. Focus
was optimized near 120 \si{nm} as that was the optimized wavelength of
the SISTINE optical design. Several line groups in Figure
\ref{fig:specfocus} were fit across various fold mirror positions to
determine best focus. For final characterization of performance, a
source such as a PtNe hollow cathode lamp, similar to those used for
wavelength calibration of HST-COS\cite{Oliveira_2010}, would have
allowed detailed characterization of resolving power over the full
bandpass. At the time of spectrograph assembly and characterization
for SISTINE it was not possible to find a vendor willing to fabricate
a PtNe hollow cathode source with an \mgf{} window. While deuterium
spectra provided examples of how well the spectrograph could perform,
it is difficult to extract a clear characterization due to the density
of lines versus the resolving power of the system. Ultimately it was
possible to demonstrate that the spectrograph had significant
performance margin and would be limited by the telescope
PSF. Characterization of resolving power was accomplished by switching
the light source to a hollow cathode lamp fed by air. Results for both
the spectrograph with a specialized calibration aperture and for the
full nominal system are summarized in Table \ref{tab:rpower}. With the
spectrograph test aperture removed the typical resolving power
observed in the ``long tank'' near 120 \si{nm} was R $\sim1500$.

\begin{table}[h!]
\caption{Resolving power measured with a hollow cathode lamp}
\label{tab:rpower}
\begin{center}
\begin{tabular}{l l l l l l l} 
 \hline
 Wavelength (\si{nm}) & 113.3 & 115.3 & 120.0 & 124.3 & 141.1 & 149.4 \\
 \hline
 Spectrograph & 4460 & 4400 & 4280 & 3950 & 4470 & 5100 \\
 Full System & 1570 & 1560 & 1490 & 1580 & 1700 & 1690 \\
 \hline
\end{tabular}
\end{center}
\end{table}

\begin{figure}[ht]
  \centering
  \includegraphics[width=\textwidth]{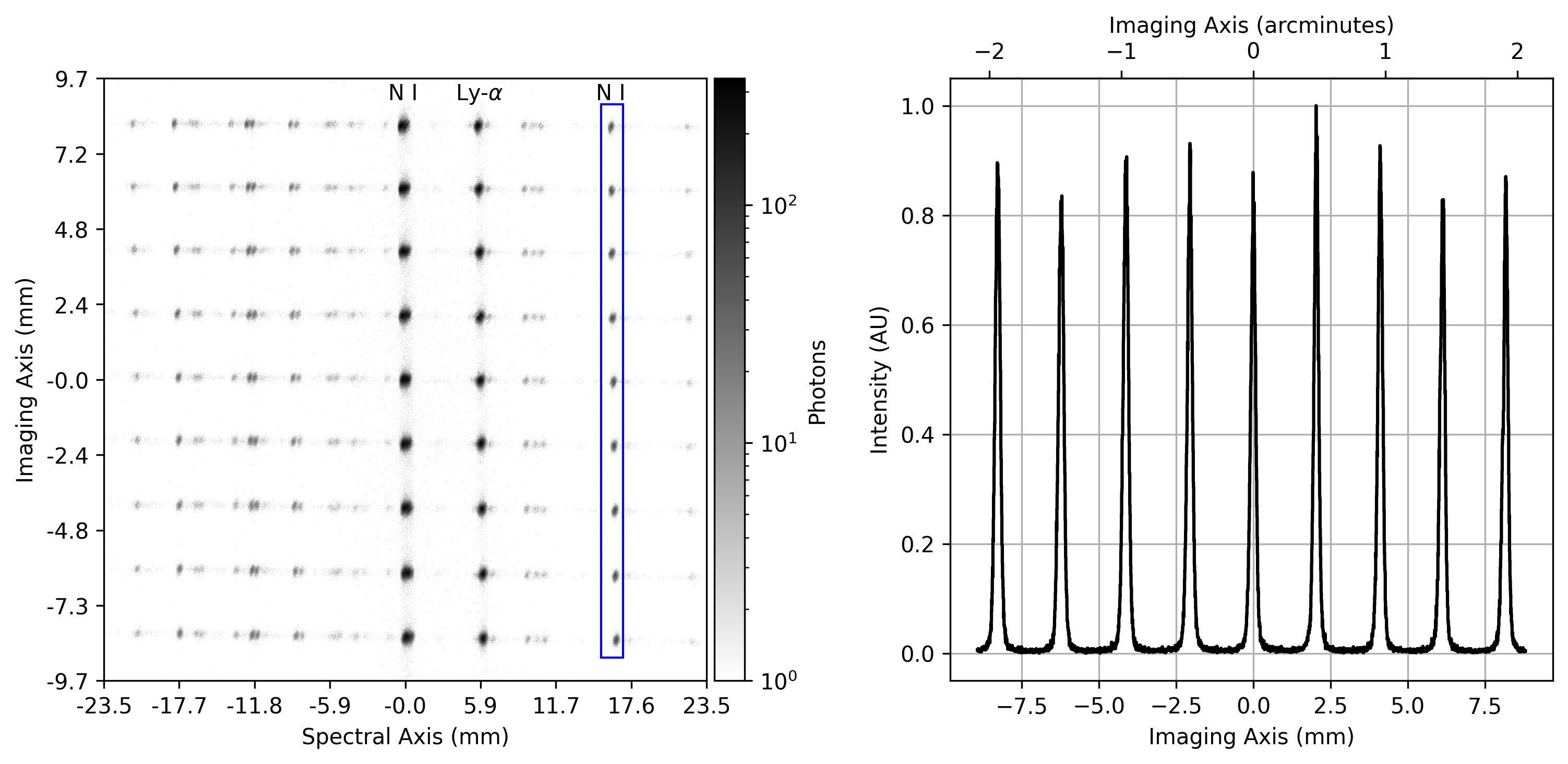}
  \caption{\label{fig:specimag} The left plot shows a portion of the
    SISTINE detector illuminated using a pinhole array to test the
    imaging capability of the instrument. The wavelength range
    displayed is roughly 114 -- 126 \si{\nm}. The data in the blue box
    (N I 124.3 \si{\nm}) is summed along the spectral axis and
    displayed in the right plot. The imaging axis maintains telescope
    limited performance across at least $\sim$\ang{;4;}.}
\end{figure}

The spectrograph focus was optimized for resolving power as it was
found that all imaging requirements were met when resolving power was
optimized. Only the first flight of SISTINE targeted an extended
object, NGC 6826, but no target data was obtained during that
flight. Subsequent flights of SISTINE observed stars and all target
stars had sufficient on-sky separation to be clearly resolved by
SISTINE in the imaging axis. Roll positions were chosen in advance to
place the long axis of the SISTINE slit along the axis of greatest
separation between stars. Figure \ref{fig:specimag} shows an imaging
axis test of SISTINE. A custom pinhole mask consisting of nine
pinholes spaced by 0.33 mm was created for the ``long tank''
collimator. These pinholes were rotationally aligned with the SISTINE
long slit axis to test imaging performance across $\sim$\ang{;4;} of
the imaging axis. The imaging axis maintains performance limited by
the combination of the telescope and ``long tank'' collimator,
$\sim$\ang{;;2.8}, across the detector extent illuminated with this
mask. This was the largest angular extent able to be tested without
mechanical modification of the ``long tank'' light source
aperture. The final PSF achieved with the SISTINE telescope sets the
upper limit on resolving power of the system and limits the ability to
reconstruct the stellar Ly$\alpha$ profile, however, sufficient
resolving power is demonstrated to resolve all other emission lines
throughout the SISTINE bandpass.

\subsection{Assembly Schedule}

The timeline for the assembly, alignment, and characterization of
SISTINE was heavily abbreviated. The first launch of SISTINE was
targeted for mid-2019 in advance of an anticipated campaign to
Australia, which at the time was scheduled for mid-2020. The SISTINE
mission initiation conference was held in March of 2018. Major
components such as the detector, optics, and skins were on site at
LASP in late 2018 and the final optical coating was completed and
measured in January of 2019. Assembly of the instrument began in late
January of 2019 and was completed by the end of June 2019. The payload
and team traveled to White Sands Missile Range on July 7, 2019 to
integrate the instrument with the NSROC payload. The entire process of
assembling, optically aligning, and characterizing SISTINE for the
first launch was accomplished in under six months. Characterization
was performed between each launch in the ``long tank'' to verify
suitable operation before shipping the instrument to the field on
subsequent launches. 

\section{Flight Performance} \label{sec:flight_performance}

SISTINE was launched three times with details summarized in Table
\ref{tab:flights}. The first launch of SISTINE was unable to acquire
data due to an experiment misalignment with the payload attitude
control system during flight. The ultimate cause of this misalignment
was found to be loosening of the primary mirror during vehicle
ascent. Minor modifications were made to the mount to better hold the
mirror in place during vibration, these modifications are summarized
in Section \ref{sec:tscope}. The second flight of SISTINE successfully
acquired data on target and these results are discussed in detail in
Ref. \citenum{CruzAguirre2023}. During the second flight of SISTINE, it
was found that the telescope thermally defocused during the flight and
caused the PSF to be larger than the slit width, limiting the spectral
resolving power of the instrument. Modifications to resolve this issue
are also discussed in Section \ref{sec:tscope}. Target data was
acquired during the third flight of SISTINE and optical system
performance during that flight was consistent with characterization
expectations from the laboratory. These results are discussed in
detail in Ref. \citenum{Behr_2023} and in Behr et al. 2024 (in
preparation) but also summarized in this work.

\begin{table}[h!]
  \caption{Summary of SISTINE launches}
  \label{tab:flights}
  \begin{center}
\begin{tabular}{l l l l} 
 \hline
 Mission Number & Launch Date & Launch Site & Target \\
 \hline\hline
 36.346 & August 11, 2019 & White Sands Missile Range & NGC 6826 \\
 36.373 & November 8, 2021 & White Sands Missile Range & Procyon A+B \\
 36.339 & Jul 6, 2022 & Arnhem Space Center & $\alpha$ Centauri A+B \\
 \hline
\end{tabular}
\end{center}
\end{table}

\begin{figure}[ht]
  \centering
  \includegraphics[width=\textwidth]{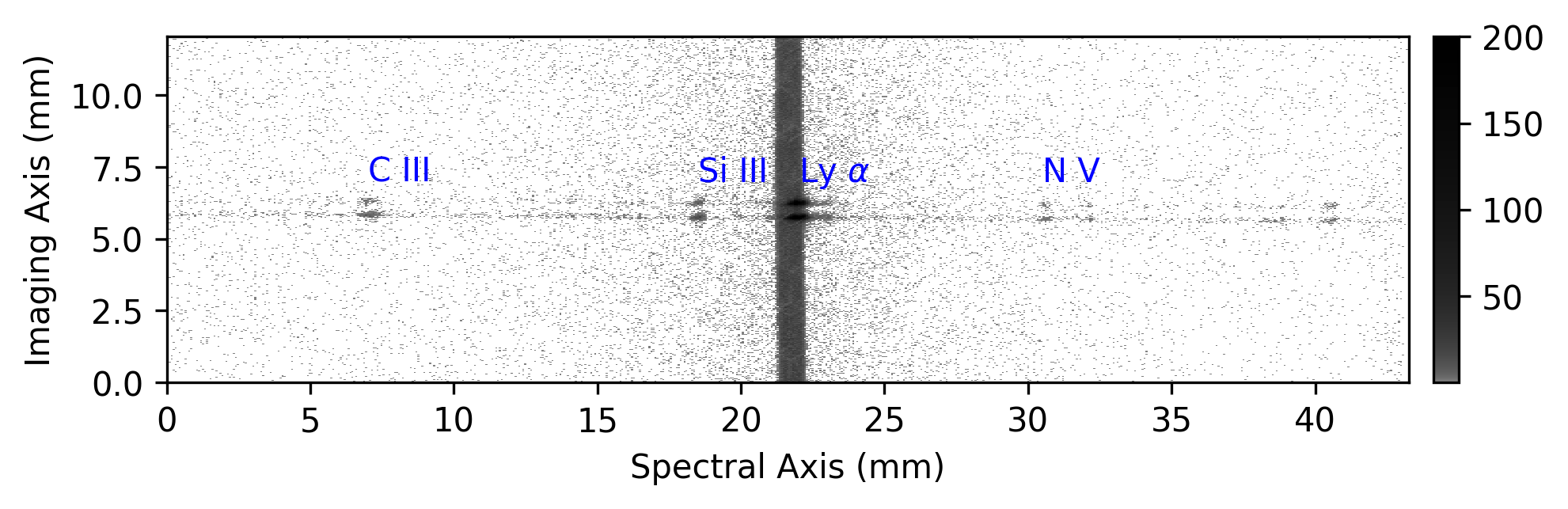}
  \caption{\label{fig:detimage} A subset of raw detector data from the
    SISTINE observation of $\alpha$ Cen A+B centered near
    Ly$\alpha$. Both stars are clearly resolved with $\alpha$ Cen B as
    the upper spectrum and $\alpha$ Cen A as the lower spectrum. The
    large vertical feature located near 22 \si{mm} on the spectral
    axis is Ly$\alpha$ airglow fully illuminating the slit. Features
    visible in the slit airglow image are MCP multifiber boundaries
    and QE grid shadows, this data has not been flat fielded to remove
    these features. The spectral axis spans from roughly 116 to 127
    \si{nm} in wavelength and the imaging axis spans approximately
    \ang{;2.8;} in angular extent. Notable spectral features are
    labeled in blue.}
\end{figure}

\begin{figure}[ht]
  \centering
  \includegraphics[width=\textwidth]{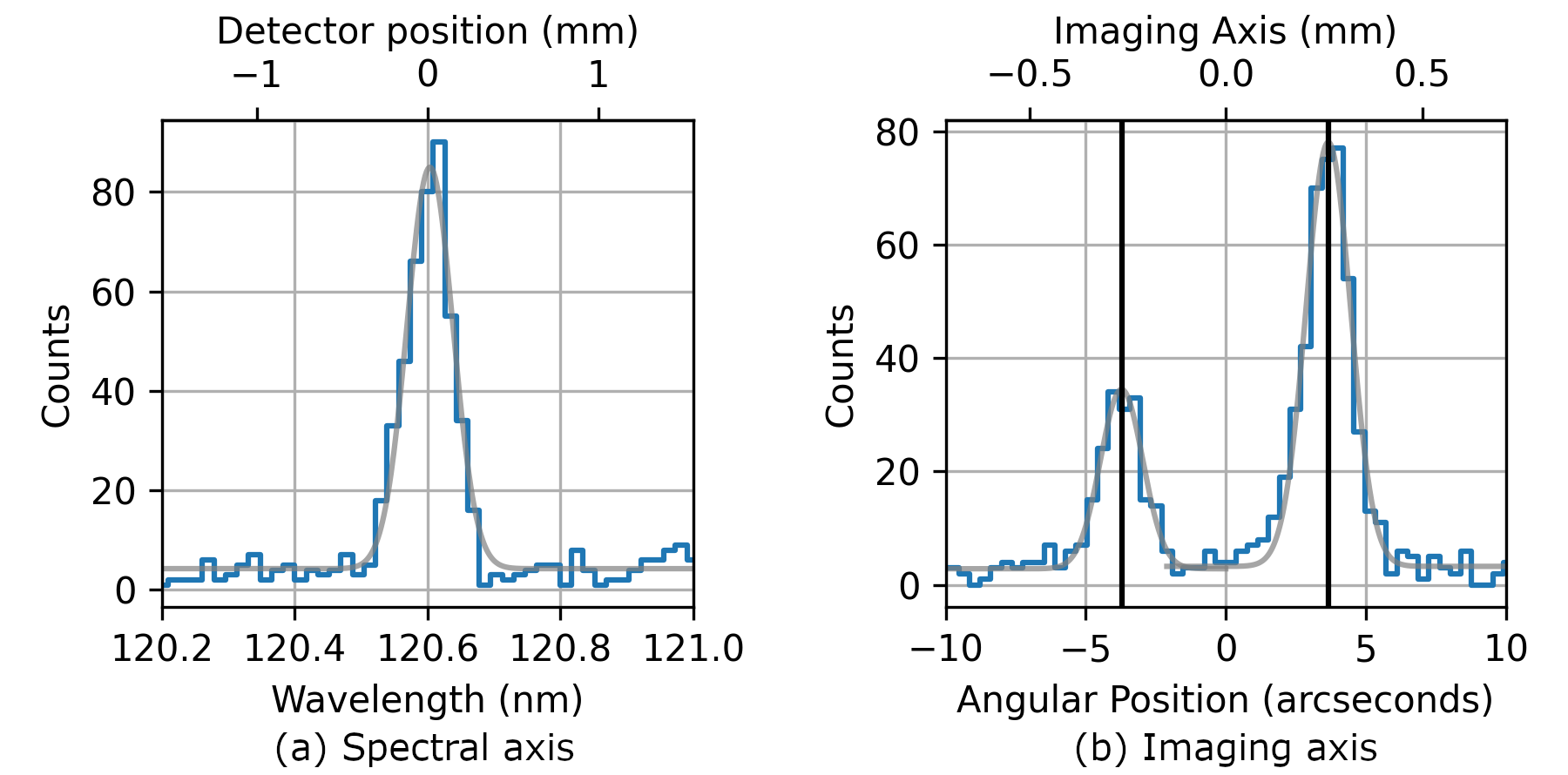}
  \caption{\label{fig:si_iii_plots} This figure shows flight data from
    SISTINE-3 in blue and Gaussian fits to this data in grey.  Panel
    (a) shows a plot of the spectrum of the Si III line for $\alpha$
    Cen A with a fit FWHM of 0.081 \si{\nm}. Panel (b) shows a one
    dimensional plot along the imaging axis of SISTINE for $\alpha$
    Cen A+B (B the left, A on the right) demonstrating that both stars
    are well resolved. Gaussian fits produce imaging axis FWHMs of
    \ang{;;1.88} and \ang{;;1.78} respectively with a separation of
    \ang{;;7.38}.}
\end{figure}

An image of a subset of the detector area centered near Ly$\alpha$
from the third flight of SISTINE is shown in Figure \ref{fig:detimage}
with imaging and spectral axes labeled. The integration time for this
image is 265 s, spanning the on-target exposure time following the
real-time pointing adjustment to place the spectrograph slit onto the
target stars. The Si III (120.6 \si{\nm}) line was used to check
instrument performance following this flight due to its substantial
signal to noise ratio and proximity to the optimized wavelength for
SISTINE (120 \si{\nm}). Figure \ref{fig:si_iii_plots} shows spectral
and imaging axis performance for the Si III line. A spectral plot for
the $\alpha$ Cen A Si III line is shown in Figure
\ref{fig:si_iii_plots}. A Gaussian fit is shown in grey with a FWHM of
0.081 \si{nm} and thus a resolving power of 1488.  Figure
\ref{fig:si_iii_plots} shows an imaging axis plot of the Si III lines
for $\alpha$ Cen A+B. Both stars have Gaussian fits shown in grey and
are clearly resolved. The FWHMs for $\alpha$ Cen A+B are \ang{;;1.88}
and \ang{;;1.78} respectively with a separation of
\ang{;;7.38}. Measurements across other points in the spectrum are
consistent with the results shown from Si III and the performance in
flight is consistent with characterization efforts from the lab.

\begin{figure}[htb]
  \centering
  \includegraphics[width=0.8\textwidth]{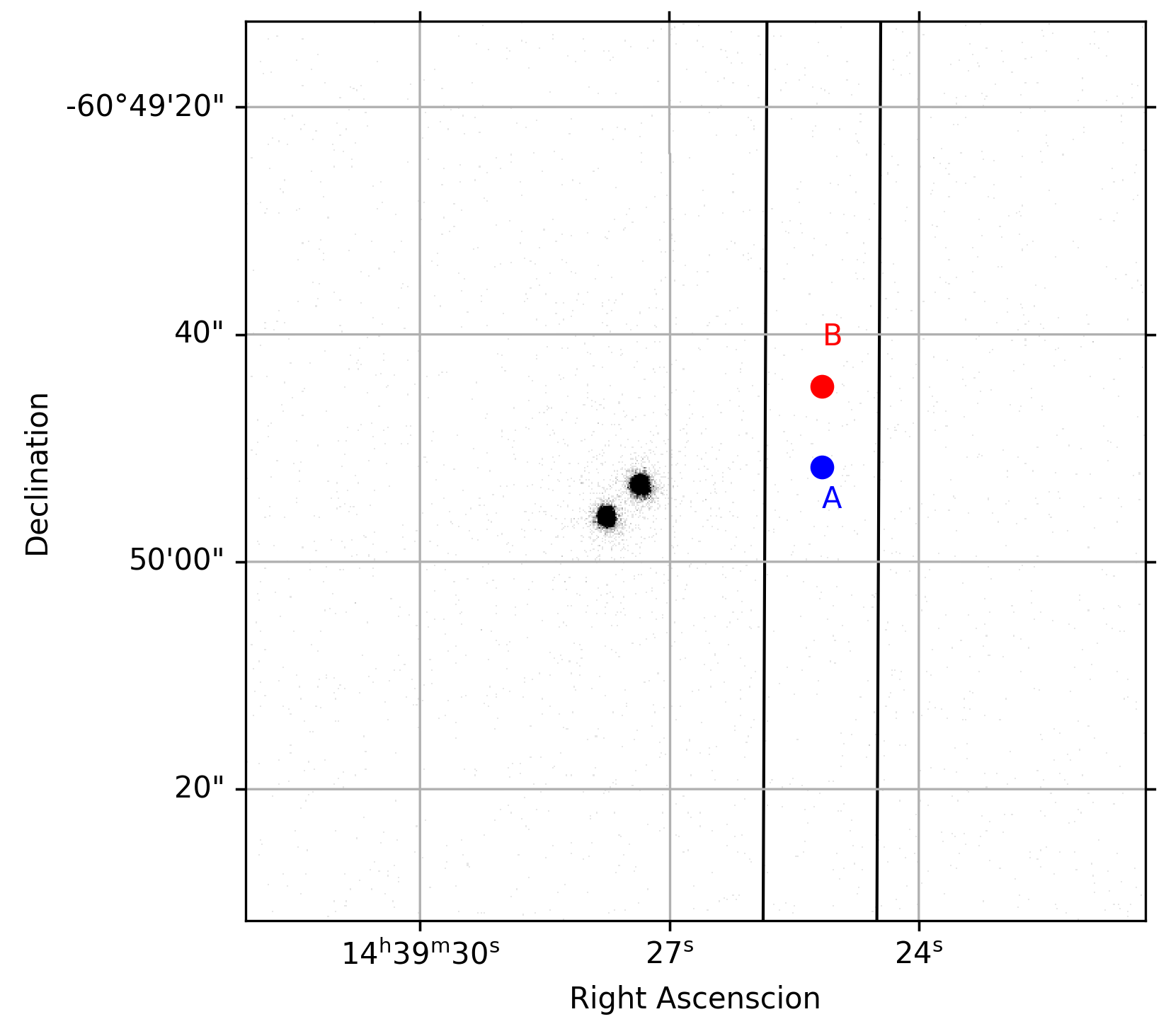}
  \caption{\label{fig:pointing} Prediction of the orientation of
    $\alpha$ Cen A+B on July 6, 2022. $\alpha$ Cen A+B is a high
    proper motion system with an orbit that changes the position angle
    on-sky appreciably over modest timescales\cite{Pourbaix_2016,
      Akeson_2021}. The desired slit orientation on sky is shown in
    black (note that the full slit length is not shown). The
    prediction is overplotted on a Chandra X-ray Observatory image
    from 2016\cite{Ayres_2014}.}
\end{figure}

Target acquisition for SISTINE 3 was particularly important due to the
orbit of $\alpha$ Cen A+B and the timescales on which those positions
on sky change. Because the slit size of SISTINE is only \ang{;;10},
both the proper motion and angle of the stars relative to the slit
need to be well understood in order to properly point the instrument
in flight to maximize separation of the stars along the imaging
axis. Observation efficiency on a sounding rocket flight is critical
due to the small amount of available observing time ($\sim$330 s for
SISTINE-3). Observation efficiency is optimized by planning a good
initial on-sky orientation to minimize pointing adjustments required
during the available observing time. Orbital parameters from
Ref. \citenum{Pourbaix_2016} were used to propagate the locations of
$\alpha$ Cen A+B to the predicted launch night time. Results of this
predicted propagation are shown in Figure \ref{fig:pointing}. This
method was verified against Chandra X-ray Observatory images of
$\alpha$ Cen A+B from 2005 and 2016\cite{Ayres_2014}. The angular
orientation in flight was consistent with the predictions. In flight,
only fine adjustments were made to the position of the slit on sky but
the instrument roll angle was not adjusted. The predicted separation
was \ang{;;7.12}, a \ang{;;0.26} difference from the measured
separation from flight data. It has come to our attention that a
recent publication with updated astrometry information was published
during the time that SISTINE-3 observations were being planned. Our
prediction for angular separation is in agreement with this
publication and we find a small difference of \ang{0.81} in position
angle\cite{Akeson_2021}.

\section{Conclusion}

SISTINE is a sounding rocket mission designed to probe the FUV
radiation environment of nearby stars. SISTINE was launched three
times with two of three missions successfully acquiring target
data. These flights successfully demonstrated the capabilities of
technologies including eLiF, ALD \alf{} protective layers, and large
format ALD processed MCPs. Although SISTINE did not meet all of the
goals originally set out, the capability of the instrument to achieve
the science goals has been shown along with the capability to handle
the launch environment. The overall design worked as intended and
would be capable of achieving higher resolving power with suitable
time investment on critical aspects of the telescope and
spectrograph. Ultimately SISTINE has proven to be a capable FUV
imaging spectrograph enabled by significant technology development
efforts that are anticipated to be significant for future UV
observatories.

\FloatBarrier

\subsection* {Code, Data, and Materials Availability} 

The data that support the findings of this paper are not publicly
available. They can be requested from the author at
\url{nicholas.nell@lasp.colorado.edu}. 

\acknowledgments 

We thank John Schwenker for inspecting the primary mirror prior to the
initial launch. The spatial testing pinholes were fabricated by
Alexandra B. Artusio-Glimpse at the National Institute of Standards
and Technology (NIST) in the Boulder Microfabrication Facility (BMF)
through funding from NIST-on-a-Chip (NOAC). We thank Michael Kaiser,
Jarrod Puseman, and Matthew Bridges for their assistance with SISTINE
assembly, testing, and support equipment. We acknowledge the hard work
and dedication of the NASA Wallops Flight Facility/NSROC payload team,
the Physical Sciences Laboratory at New Mexico State University, and
the Navy team at WSMR. We thank the NASA Sounding Rockets Program
Office, Equatorial Launch Australia, and the Gumatj Corporation for
their roles in making the 2022 launch of SISTINE from Australia
possible. SISTINE was supported by grant numbers NNX16AG28G and
80NSSC20K0412 from NASA.

\FloatBarrier


\bibliography{sisrefs}   
\bibliographystyle{spiejour}   

\end{document}